\documentclass[preprint,aps,showpacs]{revtex4}

\begin{document}

\newcommand{\dfrac}[2]{\frac{\displaystyle #1}{\displaystyle #2}}
\preprint{VPI--IPPAP--01--02}

\title{Exact Solution of the Harmonic Oscillator in Arbitrary 
Dimensions with Minimal Length Uncertainty Relations}
\author{
Lay~Nam~Chang\footnote{electronic address: laynam@vt.edu},
Djordje~Minic\footnote{electronic address: dminic@vt.edu},
Naotoshi~Okamura\footnote{electronic address: nokamura@vt.edu}, and
Tatsu~Takeuchi\footnote{electronic address: takeuchi@vt.edu}
}
\affiliation{Institute for Particle Physics and Astrophysics,
Physics Department, Virginia Tech, Blacksburg, VA 24061}

\date{November 20, 2001; revised March 15, 2002}

\begin{abstract}
We determine the energy eigenvalues and eigenfunctions of the harmonic
oscillator where the coordinates and momenta are assumed to obey the
modified commutation relations 
$[\hat{x}_i,\hat{p}_j]=i\hbar[(1+\beta\hat{p}^2)\delta_{ij}
+ \beta'\hat{p}_i\hat{p}_j]$.
These commutation relations are motivated by the fact they lead to 
the minimal length uncertainty relations which appear
in perturbative string theory.
Our solutions illustrate how certain features of string theory may 
manifest themselves in simple quantum mechanical systems through
the modification of the canonical commutation relations.
We discuss whether such effects are observable in 
precision measurements on electrons trapped in strong magnetic fields.
\end{abstract}


\maketitle

\section{Introduction}

In this paper, we derive the exact solution to the Schr\"odinger equation 
for the harmonic oscillator when the commutation relation between the
position and momentum is modified from the canonical one to 
\begin{equation}
[ \hat{x}, \hat{p} ] = i\hbar (1 + \beta \hat{p}^2) \;.
\label{Eq:Com1}
\end{equation}
This commutation relation leads to the uncertainty relation
\begin{equation}
\Delta x \ge \frac{\hbar}{2}
             \left( \frac{1}{\Delta p} + \beta\,\Delta p
             \right)\;,
\label{Uncert}
\end{equation}
which implies the existence of a minimal length
\begin{equation}
\Delta x_\mathrm{min} = \hbar\sqrt{\beta}\;,
\label{MinLength}
\end{equation}
below which the uncertainty in position, $\Delta x$, 
cannot be reduced \cite{Kempf:1995su,Kempf:1997fz}.

The minimal length uncertainty relation, Eq.~(\ref{Uncert}),
has appeared in the context of perturbative string theory \cite{gross} 
where it is a consequence of the fact that strings
cannot probe distances below the string scale $\hbar\sqrt{\beta}$.
Though the modified commutation relation, Eq.~(\ref{Eq:Com1}), 
has not been so far derived directly from string theory, 
the fact that it implies Eq.~(\ref{Uncert}) suggests that it is
one possible way in which certain features of
string theory may manifest themselves in
low energy quantum mechanical systems.

It should be noted, however, that Eq.~(\ref{Uncert}) is not an
ubiquitous prediction of string theory.
Indeed, both in the realms of perturbative and
non-perturbative string theory (where distances shorter than
the string scale can be probed by $D$-branes \cite{joep}), 
another type of uncertainty relation involving both spatial and time 
coordinates has been found to hold \cite{stu}. 
The distinction (and relation) between the minimal length uncertainty 
relation and the space-time uncertainty relation 
has been elucidated by Yoneya \cite{yoneya}.

Nontheless, Eq.~(\ref{Uncert}) does embody an intriguing 
UV/IR relation :
when $\Delta p$ is large (UV), $\Delta x$ is proportional to 
$\Delta p$ and therefore is also large (IR).  
This type of UV/IR relation has appeared
in several other contexts: the $AdS/CFT$ correspondence
\cite{adscft}, non-commutative field theory \cite{ncft},
and more recently in attempts at understanding quantum
gravity in asymptotically de Sitter spaces \cite{dscft,dsrg}.

It has been argued that the kind of
UV/IR ``bootstrap'' described by Eq.~(\ref{Uncert})
is necessary to understand observable implications of 
short distance physics on inflationary cosmology \cite{greene}.
Likewise, Banks has argued that some kind of UV/IR relation
should be at the core of the cosmological constant problem \cite{banks} 
as well as its relation to the problem of supersymmetry breaking.
Therefore, both Eq.~(\ref{Uncert}) and the underlying
Eq.~(\ref{Eq:Com1}) are well motivated
by a variety of applications, including the cosmological
constant problem which we will discuss is a subsequent paper
\cite{Chang:2001bm}.

Furthermore, the UV/IR relation
represented by Eq.~(\ref{Uncert}) suggests that certain ``stringy''
short distance (UV) effects may manifest 
themselves at longer distances (IR), lending
additional justification to our analysis of the
non-relativistic harmonic oscillator.

The problem of solving for the energy eigenvalues and eigenstates
of the harmonic oscillator with the minimal length uncertainty relation
has been studied previously by Kempf et al. in
Refs.~\cite{Kempf:1995su,Kempf:1997fz}.
However, the exact result had been obtained only for the 1-dimensional case. 
We present here the exact solution
for the general $D$-dimensional isotropic harmonic oscillator.

\section{The Harmonic Oscillator in 1 Dimension}

We represent the position and momentum operators obeying Eq.~(\ref{Eq:Com1})
in momentum space by
\begin{eqnarray}
\hat{x} & = & i\hbar\left[ (1+\beta p^2)\frac{\partial}{\partial p}
                             +\gamma\,p
                    \right]\;,  \cr
\hat{p} & = & p\;.
\label{Rep1D}
\end{eqnarray}
The choice of the constant $\gamma$ determines the
weight function in the definition of the inner product:
\begin{equation}
\langle f | g \rangle
= \int \frac{dp}{(1 + \beta p^2)^{1-\alpha}}\,f^*(p)\,g(p)\;,
\label{Product1D}
\end{equation}
where
\begin{equation}
\alpha = \frac{\gamma}{\beta}\;.
\label{alpha1D}
\end{equation}
This definition ensures the hermiticity of $\hat{x}$.
In the following, we will keep $\gamma$ arbitrary, though eventually
we will find that the energy levels in fact do not depend on its
value.

The Schr\"odinger equation for the harmonic oscillator with Hamiltonian
\begin{equation}
\hat{H} = \frac{1}{2}\mu\omega^2 \hat{x}^2 + \frac{1}{2\mu}\hat{p}^2 \;,
\end{equation}
is given by
\begin{eqnarray}
& &
\Biggl[ -\mu\hbar\omega
     \Biggl\{
          \left( (1+\beta p^2)\frac{\partial}{\partial p}
          \right)^2
        + 2\gamma p 
          \left( (1+\beta p^2)\frac{\partial}{\partial p}
          \right)
     \Biggr.
\Biggr. \cr
& & \quad
\Biggl.
     \Biggl.
        + \gamma (\beta + \gamma) p^2
        + \gamma
     \Biggr\}
     +\frac{1}{\mu\hbar\omega}\,p^2
\Biggr]\Psi(p)
= \frac{2E}{\hbar\omega}\,\Psi(p)\;.
\label{Psip}
\end{eqnarray}
A change of variable from $p$ to
\begin{equation}
\rho \equiv \frac{1}{\sqrt{\beta}} \tan^{-1}(\sqrt{\beta}p) \;,
\end{equation}
maps the region $-\infty < p < \infty$ to
\begin{equation}
-\frac{\pi}{2\sqrt{\beta}} < \rho < \frac{\pi}{2\sqrt{\beta}} \;,
\end{equation}
and casts Eq.~(\ref{Psip}) into the form:
\begin{eqnarray}
& &
\Biggl[
\mu\hbar\omega
  \Biggl\{ \frac{\partial^2}{\partial\rho^2}
        + \left( \frac{2\gamma}{\sqrt{\beta}}
                 \tan \sqrt{\beta}\rho
          \right)
          \frac{\partial}{\partial\rho}
  \Biggr\}
\Biggr. \cr
& & 
- \Biggl\{ \frac{1}{\mu\hbar\omega\beta}
        - \mu\hbar\omega\gamma
          \left( 1 + \frac{\gamma}{\beta} \right)
  \Biggr\}
  \tan^2 \sqrt{\beta}{\rho}\,
\Biggl.
+ \Biggl\{ \frac{2E}{\hbar\omega} + \mu\hbar\omega\gamma
  \Biggr\}
\Biggr]\Psi(p) = 0\;.
\end{eqnarray}
Defining dimensionless parameters by
\begin{equation}
\xi\equiv \frac{\rho}{\sqrt{\mu\hbar\omega}}\;,\quad
\kappa\equiv \sqrt{\mu\hbar\omega\beta}\;,\quad
\delta\equiv \frac{\gamma}{\beta}\;,\quad
\varepsilon\equiv\frac{2E}{\hbar\omega}\;,
\end{equation}
we obtain
\begin{equation}
\left[ \frac{\partial^2}{\partial\xi^2}
       +2\,\kappa\,\delta\,\frac{s}{c}\frac{\partial}{\partial\xi}
       -\left\{\frac{1}{\kappa^2} - \kappa^2\delta(1+\delta)
        \right\}\frac{s^2}{c^2}
       +( \varepsilon + \kappa^2\delta )
\right]\Psi(\xi) = 0\;,
\end{equation}
where we use the shorthand notations
\begin{equation}
c\equiv \cos\kappa\xi\;,\qquad
s\equiv \sin\kappa\xi\;.
\end{equation}
Let $\Psi(\xi) = c^{\lambda+\delta}\,f(s)$, where $\lambda$ is a constant to
be determined.
Then the equation for $f(s)$ is
\begin{equation}
(1-s^2)f'' - (2\lambda + 1)\,s\,f'
+ \left[ \left\{ \frac{\varepsilon}{\kappa^2} - \lambda
         \right\}
       + \left\{ \lambda(\lambda-1)-\frac{1}{\kappa^4} 
         \right\}\frac{s^2}{c^2}
  \right] f
= 0\;.
\label{Eq:f11}
\end{equation}
The variable is now $-1 < s < 1$.
Note that $\delta=\gamma/\beta$ is eliminated from the equation.
We fix $\lambda$ by requiring the coefficient of the
tangent squared term to vanish:
\begin{equation}
\lambda(\lambda-1) - \frac{1}{\kappa^4} = 0\;.
\end{equation}
The wave function should be non--singular at $c=0$, which implies
\begin{equation}
\lambda = \frac{1}{2}
        + \sqrt{ \frac{1}{4} + \frac{1}{\kappa^4} }\;.
\end{equation}
This simplifies Eq.~(\ref{Eq:f11}) to
\begin{equation}
(1-s^2)\,f'' - (\,2\lambda + 1\,)\,s\,f'
+ \left( \frac{\varepsilon}{\kappa^2} - \lambda \right) f
= 0\;.
\label{Eq:f12}
\end{equation}
Similarly, $f(s)$ should be non-singular at $s=\pm 1$.
Thus we require a polynomial solution to Eq.~(\ref{Eq:f12}).  
This requirement imposes the following condition on the coefficient of $f$:
\begin{equation}
\frac{\varepsilon}{\kappa^2} - \lambda
= n \,(\, n + 2\lambda \,) \;,
\label{Eigen1}
\end{equation}
where $n$ is a non--negative integer \cite{SpecialFunctions}.
Eq.~(\ref{Eq:f12}) becomes
\begin{equation}
(1-s^2)f'' - (\, 2\lambda + 1 \,)\,s\,f'
+ n \,(\, n + 2\lambda \,)\, f = 0\;,
\label{Eq:f13}
\end{equation}
the solution of which is given by the Gegenbauer polynomial:
\begin{equation}
f(s) = C_n^{\lambda}(s)\;.
\end{equation}
The energy eigenvalues follow from the condition Eq.~(\ref{Eigen1}):
\begin{eqnarray}
\varepsilon_n
& = & \kappa^2 \left[\, n^2 + (2n+1)\lambda \,\right]\cr
& = & \kappa^2 \left( n^2 + n + \frac{1}{2} \right)
    + \left( 2n+1 \right)
      \sqrt{ 1 + \frac{\kappa^4}{4} }\;,
\end{eqnarray}
or more explicitly,
\begin{equation}
E_n = \hbar\omega
\left[ \left( n + \frac{1}{2} \right)
       \sqrt{ 1 + \frac{\beta^2\mu^2\hbar^2\omega^2}{4} }
     + \left( n^2 + n + \frac{1}{2} \right)
       \frac{\beta\mu\hbar\omega}{2}
\right]\;.
\label{En1D}
\end{equation}
This result agrees with Kempf \cite{Kempf:1995su}.
The normalized energy eigenfunctions are:
\begin{equation}
\Psi_n(p) = 
2^{\lambda}\Gamma(\lambda)
\sqrt{ \dfrac{ n!\,(n+\lambda)\,\sqrt{\beta} }{ 2\pi\,\Gamma(n+2\lambda) } }
\;c^{\lambda + \delta}\,C_n^{\lambda}(s)\;,
\end{equation}
where
\begin{eqnarray}
c & = & \cos\sqrt{\beta}\rho \;=\; \dfrac{ 1 }{ \sqrt{1+\beta p^2} }\;,\cr
s & = & \sin\sqrt{\beta}\rho \;=\; \dfrac{ \sqrt{\beta} p } 
                                              { \sqrt{1+\beta p^2} }\;.
\end{eqnarray}

\section{The Harmonic Oscillator in $D$--Dimensions}

In more than 1 dimension,
the modified commutation relation can be generalized to
the tensorial form:
\begin{equation}
[ \hat{x}_i, \hat{p}_j ]
= i\hbar( \delta_{ij}
          + \beta \hat{p}^2 \delta_{ij}
          + \beta' \hat{p}_i \hat{p}_j
        )\;.
\label{Eq:Com2}
\end{equation}
If the components of the momentum $\hat{p}_i$ are assumed
to commute with each other, 
\begin{equation}
[ \hat{p}_i, \hat{p}_j ] = 0\;,
\end{equation}
then the commutation relations among the coordinates $\hat{x}_i$ 
are almost uniquely determined by the Jacobi Identity  
(up to possible extensions) as \cite{Kempf:1995su,Kempf:1997fz}
\begin{equation}
[ \hat{x}_i, \hat{x}_j ]
= i\hbar\,
  \frac{(2\beta-\beta')+ (2\beta+\beta')\beta \hat{p}^2 }
       { (1+\beta \hat{p}^2) }
  \left( \hat{p}_i \hat{x}_j - \hat{p}_j \hat{x}_i
  \right)\;.
\end{equation}
These operators are realized in momentum space as
\begin{eqnarray}
\hat{x}_i & = & i\hbar
\left[ (1 + \beta p^2)\frac{\partial}{\partial p_i}
       + \beta' p_i p_j \frac{\partial}{\partial p_j}
       + \gamma \,p_i
\right] \;,\cr
\hat{p}_i & = & p_i \;.
\label{RepND}
\end{eqnarray}
The arbitrary constant $\gamma$ in the representation 
of $\hat{x}_i$ does not affect the commutation relations
among the $\hat{x}_i$'s.
Again, its choice determines the weight function in the definition of the 
inner product:
\begin{equation}
\langle f | g \rangle
= \int \frac{ d^D{\mathbf{p}} }{ [1+(\beta+\beta') p^2]^{1-\alpha} }\,
       f^*({\mathbf{p}})\,g({\mathbf{p}})\;,
\label{ProductND}
\end{equation}
where
\begin{equation}
\alpha = \dfrac{ \gamma - \beta'\left(\frac{D-1}{2}\right) }
               { (\beta + \beta') } \;.
\label{alpha2D}
\end{equation}
Note that when $\beta'=0$, this expression reduces to that of the
1D case, Eq.~(\ref{alpha1D}).
Ref.~\cite{Kempf:1997fz} uses
\begin{equation}
\gamma = \beta + \beta'\left( \frac{D+1}{2} \right) \;,
\end{equation}
in which case the weight function is constant.
We will keep $\gamma$ arbitrary in our calculations.
As in the 1D case, we will find that the energy eigenvalues do not
depend on $\gamma$.

Since the $D$-dimensional harmonic oscillator 
\begin{equation}
\hat{H} = \frac{1}{2}\mu\omega^2 \hat{\mathbf x}^2
        + \frac{1}{2\mu} \hat{\mathbf p}^2
\end{equation}
is rotationally symmetric,
we can assume that the momentum space energy eigenfunctions
expressed in terms of the radial momenta can be written as
a product of spherical harmonics and a radial wave function:
\begin{equation}
\Psi_D({\mathbf p}) 
= Y_{\ell_{(D-1)}\cdots\ell_2\ell_1}(\Omega) R(p) \;.
\label{Eq:YellD}
\end{equation}
In $2D$ and $3D$, we have
\begin{eqnarray}
\Psi_{2}({\mathbf p}) 
& = & {\displaystyle \frac{1}{\sqrt{2\pi}}e^{-im\phi}\,R(p)}\;,\cr
\Psi_{3}({\mathbf p}) 
& = & Y_{\ell m}(\theta,\phi)\,R(p)\;.
\end{eqnarray}
Eq.~(\ref{Eq:YellD}) allows us to make the replacements
\begin{eqnarray}
\sum_{i=1}^N \frac{\partial^2}{\partial p_i^2}
& = & \frac{\partial^2}{\partial p^2}
    + \frac{D-1}{p}\frac{\partial}{\partial p} 
    - \frac{L^2}{p^2}\;, \cr
\sum_{i=1}^N p_i\frac{\partial}{\partial p_i} 
& = & p\frac{\partial}{\partial p}\;,
\end{eqnarray}
where \cite{Chodos:1983zi}
\begin{equation}
L^2 = \ell (\ell + D -2 )\;,\qquad
\ell = 0,1,2,\cdots\;.
\end{equation}
($\ell = |m|$ for $D=2$.)

We therefore find, not unexpectedly, that
the Schr\"odinger equation for the $D$-dimensional 
harmonic oscillator can be reduced to the 1-dimensional problem 
for the radial wave function $R(p)$.  The equation for $R(p)$ is:
\begin{eqnarray}
& &
-\mu\hbar\omega
   \Biggl[
      \left\{ [1 + (\beta + \beta') p^2]\frac{\partial}{\partial p}
      \right\}^2
   \Biggr.
     +\left\{ \frac{D-1}{p} + [\,(D-1)\beta + 2\gamma\,]\, p
      \right\}
      \left\{ [1 + (\beta + \beta') p^2]\frac{\partial}{\partial p}
      \right\} 
\cr
& & 
   \Biggl.
     -\frac{L^2}{p^2} + (\,\gamma D - 2\beta L^2\,)
     +\left\{\,\gamma\,(\,\beta D + \beta' + \gamma\,)-\beta^2 L^2\,
      \right\}\, p^2
   \Biggr]\,R(p)
+ \frac{1}{\mu\hbar\omega}\,p^2\,R(p)
= \frac{2E}{\hbar\omega}\,R(p)\;.
\cr & & 
\end{eqnarray}
Introducing the change of variable
\begin{equation}
\rho \equiv \frac{1}{\sqrt{\beta+\beta'}}
\tan^{-1}\sqrt{\beta+\beta'}\,p \;,
\end{equation}
maps the region $0<p<\infty$ to 
\begin{equation}
0<\rho<\frac{\pi}{2\sqrt{\beta+\beta'}}\;,
\end{equation}
and renders the equation into the following form:
\begin{eqnarray}
& & 
\left[ -\mu\hbar\omega
   \left\{ \frac{\partial^2}{\partial\rho^2}
         + \left( \dfrac{ (D-1)\sqrt{\beta+\beta'} } 
                        { \tan\sqrt{\beta+\beta'}\rho }
                + \dfrac{ [\, (D-1)\beta + 2\gamma \,] }
                        { \sqrt{\beta+\beta'} }
                  \tan\sqrt{\beta+\beta'}\rho
           \right)
           \frac{\partial}{\partial\rho}
   \right.
\right. \cr
& & 
   \left.
        -\dfrac{ L^2(\beta+\beta') }
               { \tan^2\sqrt{\beta+\beta'}\rho }
        + (\gamma D - 2\beta L^2)
        +\dfrac{ [\,\gamma\,(\,\beta D + \beta' + \gamma\,)\,]
                 - \beta^2 L^2 
               }
               { (\beta+\beta') }
          \tan^2\sqrt{\beta+\beta'}\rho
   \right\} \cr
& & 
\left.
 + \frac{1}{ \mu\hbar\omega(\beta+\beta') }\,\tan^2\sqrt{\beta+\beta'}\rho
\right]\,R(\rho)
= \frac{2E}{\hbar\omega}\,R(\rho)\;.
\end{eqnarray}
Defining the dimensionless parameters
\begin{equation}
\xi\equiv \frac{\rho}{\sqrt{\mu\hbar\omega}}\;,\quad
\kappa\equiv \sqrt{\mu\hbar\omega(\beta+\beta')}\;,\quad
\eta\equiv \frac{\beta}{\beta+\beta'}\;,\quad
\delta\equiv \frac{\gamma}{\beta+\beta'}\;,\quad
\varepsilon\equiv\frac{2E}{\hbar\omega}\;,
\end{equation}
we obtain
\begin{eqnarray}
& & 
\frac{1}{\kappa^2}\frac{\partial^2 R}{\partial\xi^2} 
+ \left[ (D-1) \frac{c}{s}
       + \left\{\,(D-1)\eta + 2\delta\,
         \right\}\frac{s}{c}
  \right]\frac{1}{\kappa}
         \frac{\partial R}{\partial\xi}
\cr
& &
+ \Biggl[ \left\{ \frac{\varepsilon}{\kappa^2}
                     - (2\eta-1) L^2 + \delta D
          \right\}
          - \frac{L^2}{s^2}
      + \left\{ \delta(D-1)\eta + \delta(1+\delta)
              - \eta^2 L^2 - \frac{1}{\kappa^4}
        \right\} \frac{s^2}{c^2}
  \Biggr] R
= 0\;,\cr
& &
\end{eqnarray}
where we use the shorthand notation
\begin{equation}
c = \cos\kappa\xi\;,\qquad s = \sin\kappa\xi\;,
\end{equation}
as before.
Let $R(\xi) = c^{\lambda+\delta} f(s)$.
The equation for $f(s)$ is
\begin{eqnarray}
& & (1-s^2) f'' 
+ \left[ -\left\{ 2\lambda + 1 + (D-1)(1-\eta) 
          \right\} s
         +\frac{D-1}{s}
  \right] f'
\cr
& &
+ \left[ \left\{ \frac{\varepsilon}{\kappa^2}
                - (2\eta -1) L^2 -\lambda D
         \right\}
       - \frac{L^2}{s^2}
  \right. \cr
& &
  \left.
       + \left\{ \lambda^2
               - \lambda[1+(D-1)\eta]
                 - \eta^2 L^2 - \frac{1}{\kappa^4}
         \right\}\frac{s^2}{c^2}
  \right]f = 0\;.
\end{eqnarray}
Note that as in the $1D$ case, $\delta$ is eliminated.
We choose $\lambda$ to cancel the tangent squared term:
\begin{equation}
\lambda^2 - \lambda [ 1+(D-1)\eta ]
- \eta^2 L^2 - \frac{1}{\kappa^4}
= 0\;.
\end{equation}
Taking the positive root, we obtain
\begin{equation}
\lambda =  \frac{1 + (D-1)\eta}{2}
         + \sqrt{ \frac{ \{1 + (D-1)\eta\}^2 }{ 4 }
                       + \eta^2 L^2 + \frac{1}{\kappa^4}
                }\;.
\end{equation}
The equation for $f(s)$ then simplifies to
\begin{eqnarray}
& & (1-s^2) f'' \cr
& &
+ \left[ -\left\{ 2\lambda + 1 + (D-1)(1-\eta) 
          \right\} s
         +\frac{D-1}{s} 
  \right] f' \cr
& &
+ \left[ \left\{ \frac{\varepsilon}{\kappa^2}
                -\lambda D
                -(2\eta -1) L^2 
         \right\}
       - \frac{L^2}{s^2}
  \right]f = 0\;.
\end{eqnarray}
Next, let $f(s) = s^\ell g(s)$.
This substitution eliminates the centrifugal barrier term
and gives the equation for $g(s)$:
\begin{eqnarray}
& & (1-s^2) g'' \cr
& &
+ \left[ -\{ 2\lambda + 2\ell + 1 + (D-1)(1-\eta) \} s
         +\frac{2\ell + D - 1}{s}
  \right] g' \cr
& & 
+ \left[ \frac{\varepsilon}{\kappa^2}
       - 2\eta L^2 - (2\ell+D)\lambda + \ell\{ (D-1)\eta -1 \}
  \right] g = 0\;.
\end{eqnarray}
Another change of variable
\begin{equation}
z = 2s^2 - 1
\end{equation}
maps the interval $0< s < 1$ to $-1 < z < 1$ and leads to
\begin{eqnarray}
& & (1-z^2)\frac{d^2g}{dz^2} \cr
& & + \Bigl[ (b-a) - (a+b+2) z \Bigr]\frac{dg}{dz} \cr
& & 
    + \frac{1}{4}
      \left[ \frac{\varepsilon}{\kappa^2} - 2\eta L^2
             - (2\ell+D)\lambda
             +\ell\{(D-1)\eta - 1\}
      \right]g = 0 \;,
\label{Eq:gz}
\end{eqnarray}
where
\begin{eqnarray}
a & = & \lambda  - \frac{1 + (D-1)\eta}{2}
  \;=\; \sqrt{ \frac{ \{1 + (D-1)\eta\}^2 }{ 4 }
                + \eta^2 L^2 + \frac{1}{\kappa^4} }\;, \cr
b & = & \frac{D}{2} + \ell - 1 \;.
\end{eqnarray}
Once again, we need a polynomial solution to $g(z)$ to keep the wave function
regular at $z=\pm 1$.
The condition one must impose is
\begin{equation}
\frac{1}{4}
\left[ \frac{\varepsilon}{\kappa^2} - 2\eta L^2
     - (2\ell+D)\lambda
      +\ell\{(D-1)\eta - 1\}
\right]
= n'(n' + a + b + 1) \;,
\end{equation}
where $n'$ is a non--negative integer \cite{SpecialFunctions}.
This casts Eq.~(\ref{Eq:gz}) into the form
\begin{equation}
(1-z^2)\frac{d^2g}{dz^2}
+ \Bigl[ (b-a) - (a+b+2) z \Bigr]\frac{dg}{dz}
+ n'(n' + a + b + 1) g = 0 \;,
\end{equation}
the solution of which is given by the Jacobi polynomial:
\begin{equation}
g(z) = P_{n'}^{(a,b)}(z)\;.
\end{equation}
The energy eigenvalues are given by
\begin{eqnarray}
\frac{\varepsilon}{\kappa^2}
& = & 2 \left[ (2n' + \ell) + \frac{D}{2} \right] \lambda 
     +  \left[ 4{n'}^2 + 2n'(1-\eta)(D-1) + 4n'\ell + 2\eta L^2
              +\{\eta(D-1)-1\}\ell
         \right] \cr
& = & 2\left( n + \frac{D}{2} \right)\lambda
      + \left[ n^2 + n(1-\eta)(D-1) + (2\eta-1)L^2 \right] \cr
& = & 2\left( n + \frac{D}{2} \right)
       \sqrt{ \frac{ \{1 + (D-1)\eta\}^2 }{ 4 }
              + \eta^2 L^2 + \frac{1}{\kappa^4} } 
\cr & & \qquad\qquad\qquad
      + \left[ n^2 + nD + \frac{D^2}{2} \eta + \frac{D}{2} (1-\eta) 
            + (2\eta-1)L^2
        \right] \;,
\end{eqnarray}
where in the second line we have made the replacement $2n'+\ell = n$.
The final expression is:
\begin{eqnarray}
E_{n\ell}
& = & \hbar\omega
\left[
  \left( n + \frac{D}{2} \right) 
        \sqrt{ 1 + \left\{ \beta^2 L^2 
                         + \frac{ (D\beta + \beta')^2 }{ 4 }
                   \right\}\mu^2\hbar^2\omega^2
             }
\right. \cr
& & \qquad\left.
+ \left\{ (\beta + \beta')\left( n + \frac{D}{2}\right)^2
        + (\beta - \beta')\left(L^2 + \frac{D^2}{4}\right)
        + \beta'\frac{D}{2}
  \right\}\frac{\mu\hbar\omega}{2}
\right]\;.
\label{Enell}
\end{eqnarray}
This equation represents the main result of our paper.
The $1D$ result can be reproduced from this expression by setting
$D=1$, $L^2=0$, and $\beta'=0$.
The normalized energy eigenfunctions are
\begin{equation}
R_{n\ell}(p) = 
\sqrt{ \dfrac{ 2 (2n'+a+b+1)\, n'!\, \Gamma(n'+a+b+1) }
             { \Gamma(n'+a+1) \Gamma(n'+b+1) }
     }\,
(\beta+\beta')^{D/4}\, c^{\lambda+\delta}\, s^\ell\, P_{n'}^{(a,b)}(z)\;,
\end{equation}
where $n'=(n-\ell)/2$, and
\begin{eqnarray}
c & = & \cos\sqrt{\beta+\beta'}\,\rho
  \;=\; \dfrac{1}{ \sqrt{ 1 + (\beta+\beta')p^2 } }\;, \cr
s & = & \sin\sqrt{\beta+\beta'}\,\rho
  \;=\; \dfrac{ \sqrt{\beta + \beta'}\,p }{ \sqrt{1 + (\beta+\beta')p^2} }\;,\cr
z & = & 2s^2 -1
  \;=\; \dfrac{ (\beta+\beta')p^2 -1 }{ (\beta+\beta')p^2 + 1 }\;.
\end{eqnarray}

\section{Comparison with Previous Results}

Kempf \cite{Kempf:1997fz} has calculated the energy eigenvalues
of the $2D$ and $3D$ harmonic oscillators to
linear order in $\beta$ and $\beta'$.
From the exact expression Eq.~(\ref{Enell}), 
we can easily identify the terms to that order to be:
\begin{equation}
E_{n\ell}
\approx \hbar\omega
\left[
  \left( n + \frac{D}{2} \right) 
+ \frac{1}{2}
  \left\{ (k^2 + {k'}^2 )\left( n + \frac{D}{2}\right)^2
        + (k^2 - {k'}^2 )\left(L^2 + \frac{D^2}{4}\right)
        + {k'}^2\frac{D}{2}
  \right\}
\right]\;.
\end{equation}
For ease of comparison, we have introduced the notation
\begin{equation}
k^2 = \beta\mu\hbar\omega\;,\qquad
{k'}^2 = \beta'\mu\hbar\omega\;.
\end{equation}
For the $D=2$ and $D=3$ cases, the explicit expressions are
\begin{eqnarray}
E_{2D}
& \approx & 
\hbar\omega
\left[ (n+1)
+ \frac{1}{2}
  \left\{ (k^2 + {k'}^2) (n+1)^2
        + (k^2 - {k'}^2) (\ell^2 +1)
        + {k'}^2
  \right\}
\right]\;, \cr
E_{3D}
& \approx & 
\hbar\omega
\left[ \left(n+\frac{3}{2}\right)
+ \frac{1}{2}
  \left\{ (k^2 + {k'}^2) \left(n+\frac{3}{2}\right)^2
        + (k^2 - {k'}^2) \left(\ell(\ell+1) +\frac{9}{4}\right)
        + {k'}^2\frac{3}{2}
  \right\}
\right]\;. \cr
& &
\end{eqnarray}
We define the parameter $s$ by
\begin{equation}  
s \equiv n'+ 1 = \frac{n-\ell}{2}+1\;,
\end{equation}
which takes values from $1$ to $[(n+2)/2]$ for fixed $n$. Then
\begin{eqnarray}
E_{2D}
& \approx & \hbar\omega
\left[ (n+1)
+ k^2 (n^2 + 3n + 3) 
-{k'}^2 \left( n+\frac{3}{2} \right)
- 2\, (k^2 - {k'}^2) \,s\,(n+2-s)\,
\right]\;,\cr
E_{3D}
& \approx & \hbar\omega
\left[ \left(n+\frac{3}{2}\right)
+ k^2 \left(n^2 + 4n + \frac{21}{4}\right) 
-{k'}^2 \left( n+\frac{9}{4} \right)
- (k^2 - {k'}^2) \,s\,(2n+5-2s)\,
\right]\;,\cr
& &
\end{eqnarray}
which agree with Ref.~\cite{Kempf:1997fz}.
See also Ref.~\cite{Brau:1999uv}.

\section{Electrons in a Penning Trap}

The $n^2$ dependence of the harmonic oscillator energy levels, 
Eq.~(\ref{Enell}),
would be an unmistakeable signature of the modified commutation relations
Eq.~(\ref{Eq:Com1}) or (\ref{Eq:Com2}).
Even if $\beta$ and $\beta'$ were small, the deviation from the
usual $n$ dependence should be manifest for sufficiently large $n$.

The cyclotron motion of an electron in a Penning trap \cite{Mittleman:it}
is effectively a one-dimensional harmonic oscillator.
By looking for shifts in its energy levels,
we may be able to place a constraint on $\beta$.
To leading order in $\beta$ and $n$, the deviation of the harmonic 
oscillator energy levels from the canonical 
$\hbar\omega(n+\frac{1}{2})$ is given by
\begin{equation}
\frac{\Delta E_n}{\hbar\omega} 
= \left(\frac{\beta m \hbar \omega}{2}\right) n^2\;,
\end{equation}
which grows quickly with $n$. Note that the combination of parameters
that is constained by a measurement of $\Delta E_n/\hbar\omega$
is $\beta m\hbar\omega$.
The cyclotron frequency of an electron trapped
in a magnetic field of strength $B$ is (in SI units) 
\begin{equation}
\omega_c = \frac{e B}{m_e}\;.
\end{equation}
Therefore, 
\begin{equation}
m_e\hbar\omega_c = (e\hbar) B
= (1.7\times 10^{-53}\,\mathrm{kg^2\cdot m^2/s^2/T})\,B\;,
\end{equation}
which is independent of the electron mass $m_e$.
For a trapping field strength of $B = 6\,\mathrm{T}$, we obtain
\begin{equation}
m_e\hbar\omega_c = e\hbar B
= (1.0\times 10^{-52}\,\mathrm{kg^2\cdot m^2/s^2})\;.
\end{equation}
Even though we anticipate that measuring the energy levels accurately for
very large $n$ would be difficult, let us assume for the sake
of argument that deviations as large as $\hbar\omega_c$ 
would be detectable.
Then, the absence of such a deviation for the $n$-th energy level
would imply
\begin{equation}
\left( \frac{\beta e\hbar B}{2}\right) n^2 < 1\;,
\end{equation}
or
\begin{equation}
\beta < \left( \frac{2}{e\hbar B} \right) \frac{1}{n^2}
= \frac{ (2.0\times 10^{52}\,\mathrm{m^2/J^2\cdot s^2}) }
       { n^2 }\;.
\end{equation}
This translates into 
\begin{equation}
\hbar\sqrt{\beta}
< \sqrt{ \left( \frac{2\hbar}{e B} \right) \frac{1}{n^2} }
= \frac{ (1.5\times 10^{-8}\,\mathrm{m}) }{ n }\;,
\end{equation}
as a limit for the minimal length, Eq.~(\ref{MinLength}), and 
\begin{equation}
\frac{1}{\sqrt{\beta}}
> \sqrt{ \left(\frac{e\hbar B}{ 2 }\right) n^2 }
= (7.1\times 10^{-27}\,\mathrm{kg\cdot m/s})\,n
= (13\,\mathrm{eV/c})\,n\;,
\end{equation}
for the string momentum scale.

Aside from the technical question of whether one can measure the
energy levels accurately for large $n$, we must require that $n$ satisfy
\begin{equation}
\frac{n\,\hbar\omega_c}{m_e c^2} \ll 1\;,
\end{equation}
for our electron to stay non-relativistic. 
This leads to the constraint
\begin{equation}
n \ll \frac{m_e c^2}{\hbar\omega_c} 
    = \frac{(m_e c)^2}{(e\hbar)B} \approx 10^9\;.
\end{equation}
This condition also maintains the radius of the
electron's cyclotron motion to be well within the geometry of the 
Penning trap. 
Therefore, the maximum value of $n$ that can be used to
constrain $\beta$ would be $n\sim 10^8$ if we allow for a
10\% relativistic correction.
So the best limit on $\beta$ that can be achieved will be
\begin{equation}
\hbar\sqrt{\beta} < 10^{-16}\,\mathrm{m}\;,\qquad
\frac{1}{\sqrt{\beta}} > 1\,\mathrm{GeV/c}\;.
\end{equation}
Obtaining a better limit would be difficult since improving the
limit on $\sqrt{\beta}$ by an order of magnitude requires the
improvement of the limit on $\beta$ by two orders of magnitude.

\section{Discussion and Conclusions}

We have obtained the exact energy eigenvalues and eigenstates
of the harmonic oscillator when the coordinate and momentum operators
satisfy the modified commutation relations
Eq.~(\ref{Eq:Com1}) or (\ref{Eq:Com2}).

The energy levels, Eqs.~(\ref{En1D}) and (\ref{Enell}), 
grow as $n^2$ for large $n$.
The reason for this $n^2$ behavior can be understood as follows:
The change of variable from $p$ to $\rho$ in the $1D$ problem
changes the $p^2$ kinetic term into a $\tan^2\sqrt{\beta}\rho$ potential
which is bounded at $\rho = \pm\pi/2\sqrt{\beta}$.
For higher dimensions, 
the effective potential is $\tan^2\sqrt{\beta+\beta'}\rho$
plus a centrifugal barrier $\cot^2\sqrt{\beta+\beta'}\rho$
which introduces a wall at $\rho = \pi/2\sqrt{\beta+\beta'}$
in addition to the one at $\rho=0$.
For higher energy eigenstates, the potentials are in essence 
square wells, leading to the $n^2$ dependence of the energy.
Indeed, the energy eigenvalues of a spherically symmetric
square well potential of radius $\pi/2\sqrt{\beta+\beta'}$
are given approximately by
\begin{equation}
E_n \approx \hbar\omega\left(\frac{\mu\hbar\omega}{2}\right)
            \left( \beta + \beta' \right)
            \left( n + \frac{D+1}{2} \right)^2\;.
\end{equation}

The parameter $\gamma$, introduced in Eqs.~(\ref{Rep1D}) and
(\ref{RepND}) has no effect on the energy eigenvalues and
only results in the wave functions acquiring an extra factor of
\begin{equation}
[ 1+(\beta+\beta')p^2 ]^{-\delta/2}
\end{equation}
which cancels the $\gamma$ dependence in the weight function 
of the inner product, Eqs.~(\ref{Product1D}) and (\ref{ProductND}).

The original
\begin{equation}
\frac{ (D+n-1)! }{ (D-1)!\,n! } 
\end{equation}
fold degeneracy of the $n$-th energy state is broken, leaving only the
\begin{equation}
  \frac{ (D+\ell-1)! }{ (D-1)!\,\ell! } 
- \frac{ (D+\ell-3)! }{ (D-1)!\,(\ell-2)! } 
\end{equation}
fold degeneracy for each value of $\ell$ due to rotational symmetry
\cite{Chodos:1983zi}.
This loss of degeneracy can be interpreted as the breaking of
self--supersymmetry of the harmonic oscillator \cite{Cooper:dm}.
The natural question arises whether an analogue exists on 
the level of field theory as a potentially
new mechanism for supersymmetry breaking.

Potential constraints on $\beta$ that can be placed by measuring the
energy levels of an electron trapped in a strong magnetic field
have been discussed.   Even under optimistic assumptions, the constraints
that can be imposed are weak.

In addition to affecting the energy levels of quantum mechanical systems, 
the modified commutation relations,
Eqs.~(\ref{Eq:Com1}) and (\ref{Eq:Com2}), may have other far reaching
consequences.
In subsequent papers, we will discuss their effects on the
calculation of the cosmological constant \cite{Chang:2001bm},
and the motion of macrosopic objects \cite{BCMOST:2002}.

\section*{Acknowledgments}

We would like to thank Vijay Balasubramanian, Atsushi Higuchi, Asad Naqvi, 
Koenraad Schalm, Gary Shiu, Joseph Slawny,
and Matthew Strassler for helpful discussions.
This research is supported in part by a grant from the US 
Department of Energy, DE--FG05--92ER40709.



\begin{thebibliography}{99}

\bibitem{Kempf:1995su}
A.~Kempf, G.~Mangano and R.~B.~Mann,
Phys.\ Rev.\ D {\bf 52}, 1108 (1995)
[arXiv:hep-th/9412167].

\bibitem{Kempf:1997fz}
A.~Kempf,
J.\ Phys.\ A {\bf 30}, 2093 (1997)
[arXiv:hep-th/9604045].

\bibitem{gross} 
D.~J.~Gross and P.~F.~Mende,
Nucl.\ Phys.\ B {\bf 303}, 407 (1988); 
D.~J.~Gross and P.~F.~Mende,
Phys.\ Lett.\ B {\bf 197}, 129 (1987); 
D.~Amati, M.~Ciafaloni and G.~Veneziano,
Phys.\ Lett.\ B {\bf 216}, 41 (1989);
D.~Amati, M.~Ciafaloni and G.~Veneziano,
Int.\ J.\ Mod.\ Phys.\ A {\bf 3}, 1615 (1988);
D.~Amati, M.~Ciafaloni and G.~Veneziano,
Phys.\ Lett.\ B {\bf 197}, 81 (1987);
E.~Witten,
Phys.\ Today {\bf 49} (1997) 24.

\bibitem{joep}
J.~Polchinski,
Phys.\ Rev.\ Lett.\  {\bf 75}, 4724 (1995) [arXiv:hep-th/9510017]; 
M.~R.~Douglas, D.~Kabat, P.~Pouliot and S.~H.~Shenker,
Nucl.\ Phys.\ B {\bf 485}, 85 (1997) [arXiv:hep-th/9608024].

\bibitem{stu} 
T.~Yoneya,
Int.\ J.\ Mod.\ Phys.\ A {\bf 16}, 945 (2001)
[arXiv:hep-th/0010172]; 
T.~Yoneya,
arXiv:hep-th/9707002; 
M.~Li and T.~Yoneya,
Phys.\ Rev.\ Lett.\  {\bf 78}, 1219 (1997) [arXiv:hep-th/9611072];
M.~Li and T.~Yoneya,
arXiv:hep-th/9806240; 
H.~Awata, M.~Li, D.~Minic and T.~Yoneya,
JHEP {\bf 0102}, 013 (2001) [arXiv:hep-th/9906248]; 
D.~Minic,
Phys.\ Lett.\ B {\bf 442}, 102 (1998) [arXiv:hep-th/9808035].

\bibitem{yoneya}T.~Yoneya,
Prog.\ Theor.\ Phys.\  {\bf 103}, 1081 (2000)
[arXiv:hep-th/0004074].

\bibitem{adscft} 
L.~Susskind and E.~Witten,
arXiv:hep-th/9805114; 
A.~W.~Peet and J.~Polchinski, 
Phys.\ Rev. \ D {\bf 59} 065011 (1999).

\bibitem{ncft} 
M.~R.~Douglas and N.~A.~Nekrasov,
arXiv:hep-th/0106048.

\bibitem{dscft}
C.~M.~Hull,
JHEP {\bf 9807}, 021 (1998) [arXiv:hep-th/9806146];
C.~M.~Hull,
JHEP{\bf 9811}, 017 (1998) [arXiv:hep-th/9807127];
V.~Balasubramanian, P.~Horava and D.~Minic,
JHEP {\bf 0105}, 043 (2001) [arXiv:hep-th/0103171];
E. Witten, 
arXiv:hep-th/0106109;
A.~Strominger,
arXiv:hep-th/0106113.

\bibitem{dsrg} 
A. Strominger, 
arXiv:hep-th/0110087; 
V.~Balasubramanian, J.~de Boer and D.~Minic,
arXiv:hep-th/0110108.

\bibitem{greene}
A.~Kempf,
Phys.\ Rev.\ D {\bf 63}, 083514 (2001) [arXiv:astro-ph/0009209];
A.~Kempf and J.~C.~Niemeyer,
Phys.\ Rev.\ D {\bf 64}, 103501 (2001) [arXiv:astro-ph/0103225];
R.~Easther, B.~R.~Greene, W.~H.~Kinney and G.~Shiu,
Phys.\ Rev.\ D {\bf 64}, 103502 (2001) [arXiv:hep-th/0104102];
R.~Easther, B.~R.~Greene, W.~H.~Kinney and G.~Shiu,
arXiv:hep-th/0110226.

\bibitem{banks} 
T.~Banks,
Int.\ J.\ Mod.\ Phys.\ A {\bf 16} (2001) 910.
T.~Banks,
arXiv:hep-th/0007146.
For other closely related attempts to understand the cosmological
constant problem consult for example:
T.~Banks,
arXiv:hep-th/9601151; 
A.~G.~Cohen, D.~B.~Kaplan and A.~E.~Nelson,
Phys.\ Rev.\ Lett.\  {\bf 82}, 4971 (1999)
[arXiv:hep-th/9803132]; 
P.~Horava and D.~Minic,
Phys.\ Rev.\ Lett.\  {\bf 85}, 1610 (2000)
[arXiv:hep-th/0001145]; 
N.~Arkani-Hamed, S.~Dimopoulos, N.~Kaloper and R.~Sundrum,
Phys.\ Lett.\ B {\bf 480}, 193 (2000)
[arXiv:hep-th/0001197]; S.~Kachru, M.~Schulz and E.~Silverstein,
Phys.\ Rev.\ D {\bf 62}, 045021 (2000)
[arXiv:hep-th/0001206].

\bibitem{Chang:2001bm}
L.~N.~Chang, D.~Minic, N.~Okamura and T.~Takeuchi,
arXiv:hep-th/0201017.

\bibitem{Chodos:1983zi}
A.~Chodos and E.~Myers,
Annals Phys.\  {\bf 156}, 412 (1984);
A.~Higuchi,
J.\ Math.\ Phys.\  {\bf 28}, 1553 (1987);
N.~IA.~Vilenkin,
``Special Functions and the Theory of Group Representations''
(AMS 1968).

\bibitem{SpecialFunctions}
I.~S.~Gradshteyn and I.~M.~Ryzhik,
``Table of Integrals, Series and Products'' (Acedemic Press);
M.~Abramowitz and I.~A.~Stegun,
``Handbook of Mathematical Functions, with Formulas, Graphs, and
Mathematical Table'' (Dover).

\bibitem{Brau:1999uv}
F.~Brau,
J.\ Phys.\ A {\bf 32}, 7691 (1999)
[arXiv:quant-ph/9905033].

\bibitem{Mittleman:it}
L.~S.~Brown and G.~Gabrielse,
Rev.\ Mod.\ Phys.\  {\bf 58}, 233 (1986);
R.~K.~Mittleman, I.~I.~Ioannou, H.~G.~Dehmelt and N.~Russell,
Phys.\ Rev.\ Lett.\  {\bf 83}, 2116 (1999);
H.~Dehmelt, R.~Mittleman, R.~S.~van Dyck and P.~Schwinberg,
Phys.\ Rev.\ Lett.\  {\bf 83}, 4694 (1999)
[arXiv:hep-ph/9906262].

\bibitem{Cooper:dm}
F.~Cooper and B.~Freedman,
Annals Phys.\  {\bf 146}, 262 (1983).

\bibitem{BCMOST:2002}
S.~Benczik, L.~N.~Chang, D.~Minic, N.~Okamura,
S.~Rayyan, and T.~Takeuchi, in preparation.


\end{thebibliography}
\end{document}